\documentclass[aps,pra,twocolumn]{revtex4}

\usepackage{graphicx}% Include figure files

\usepackage{bbold}

\usepackage{dcolumn}% Align table columns on decimal point

\usepackage{bm}% bold math

\newcommand{\epsv}{\mbox{\boldmath$\epsilon$}}
\newcommand{\ket}[1]{|\, #1 \,\rangle}

\newcommand{\ibrkt}[2]{|\, #1 \,\rangle\langle\, #2 \,|}
\newcommand{\Tr}{{\rm Tr}}

\newcommand{\Pj}{{\cal P}}

\newcommand{\iu}{{\rm i}}
\newcommand{\uop}{\mathbb{1}}
\begin{document}

\title{Resonance fluorescence of a trapped three-level atom}
\author{Marc Bienert}\author{Wolfgang
  Merkel} \author{Giovanna Morigi}
\affiliation{Abteilung f\"ur Quantenphysik,
  Universit\"at Ulm, Albert-Einstein-Allee 11, D-89069 Ulm, Germany} 
\date{\today} 
\begin{abstract} 
\noindent 
We investigate theoretically the spectrum of resonance fluorescence 
of a harmonically trapped atom, whose internal transitions are 
$\Lambda$--shaped and 
driven at two-photon resonance by a pair of lasers, which cool the
center--of--mass motion. 
For this configuration, photons are
scattered only due to the mechanical effects of the quantum 
interaction between light
and atom. We study the spectrum of emission in the final stage of
laser--cooling, when the atomic center-of-mass dynamics is quantum mechanical
and the size of the wave packet is much smaller than the laser wavelength (Lamb--Dicke limit).
We use the spectral decomposition of the Liouville operator 
of the master equation for the atomic density matrix and apply second order
perturbation theory.  
We find that the spectrum of resonance fluorescence
is composed by two narrow sidebands ---
the Stokes and anti-Stokes components of the scattered light --- while all
other signals are in general orders of magnitude smaller. For very low
temperatures, however, 
the Mollow--type inelastic component of the spectrum
becomes visible. This exhibits novel features which allow further insight 
into the quantum dynamics of the system.
We provide a physical model that interprets our results and
discuss how one can recover temperature and cooling rate of the atom from the
spectrum. The behaviour of the considered system is
compared with the resonance fluorescence of a trapped atom whose internal
transition consists of two-levels.  
\end{abstract} 
\maketitle
\section{INTRODUCTION} 
\noindent 
Cold trapped atoms are an ideal system for investigating the quantum
properties of the mechanical effects of photon-atom 
interaction. Pioneering experiments with optical lattices and ion traps have
allowed to measure and characterize several properties of the  
scattered radiation, thereby gaining insight into the dynamics of the driven
atoms and, in particular, 
of the mechanical effects of light on the atomic center-of-mass
motion~\cite{Grynberg2001,Leibfried03}.  
Recently, in experiments with single trapped ions it has been possible to 
measure with high precision the elastic component of the light scattered
by a driven dipole~\cite{Hoffges97,Raab2000}, 
and to observe and characterize the Stokes and anti-Stokes components
due to the harmonic motion in the trap~\cite{Raab2000}, 
thereby confirming theoretical
predictions~\cite{Lindberg86,Cirac93}. 
Lately, the properties and the manifestation of the mechanical effects 
in the light scattered by these systems is
experiencing renewed interest in several experiments, investigating 
the coupling of radiation with single atoms and ions in
optical resonators~\cite{MPQ,Caltech,Eschner2001,Walther2001,Mundt2002,Kruse2003,Kuhn2002,Eschner2002}.    

In this work, we investigate the spectrum of resonance fluorescence of  
a harmonically trapped atom, whose internal transition is $\Lambda$--shaped,
and which is cooled by two lasers tuned at two-photon resonance. 
This configuration 
is peculiar, since photon emission arises only due to the 
mechanical effects in the photon-atom interaction. In fact,
when the coupling between internal and external degrees of freedom can be
neglected (e.g.\ for copropagating laser beams), this system exhibits coherent
population 
trapping~\cite{CPT,Janik85,Stalgies98}: The electronic stationary state 
is a 
stable coherence, that does not absorb photons due to destructive interference
between the dipole excitation paths, leading to no emission of photons
at steady state.
In contrast, for laser configurations where two photon processes are
Doppler-sensitive, 
internal and external degrees of freedom are coupled. If the atomic
center-of-mass motion is confined in a steep
potential, such configuration may allow for laser--cooling to the 
potential ground state~\cite{EITcooling,Giovanna03}. Here, we study the 
spectral properties of the radiation scattered by the atoms in the final stage
of the laser--cooling dynamics.

Our theoretical analysis considers the quantum dynamics of the internal and
external degrees of freedom of the atom. It
is based on the perturbative expansion of the
atomic dynamics in second order in the Lamb--Dicke parameter, i.e.\ in the
ratio between the size of the wave packet over the laser
wavelength~\cite{Stenholm86}. We extend previous theoretical
investigations~\cite{Lindberg86,Cirac93}, which analysed the Stokes and
anti-Stokes components of the radiation scattered by a trapped dipole. In
those studies these components dominate over the Mollow inelastic
spectrum of the bare dipole~\cite{Mollow}, which is mainly due to
photon scattering at zero order in the
Lamb--Dicke expansion. In our case, the spectral component of the bare
three-level atom disappears due to destructive quantum interference,
and additional features emerge, which allow further
insight into the coupled dynamics between the internal and external 
degrees of freedom of the driven
atom. We analyse each spectral component, and discuss
how to extract information from these results
about the atomic dynamics at steady state.  

This work is organized as follows. In section II the system is described and
the dynamics is discussed qualitatively. 
In section III we present the theoretical description and evaluate 
the spectrum using the formalism of~\cite{Lindberg86,Cirac93}. We apply
perturbation theory combined with the spectral decomposition of the Liouville
operator~\cite{Briegel93,Barnett2000}, and calculate contributions to
the spectrum at higher orders in the Lamb--Dicke expansion. In section 
IV the results are summarized and compared with the picture for the dynamics
presented in section II. Finally, we discuss our results in comparison 
with the spectrum of a two-level transition driven by a plane and by a 
standing wave, as evaluated in~\cite{Lindberg86,Cirac93}. 
In the Appendices, several details of the theoretical
derivation are reported. The reader who is not interested in the theoretical
details can discard section III without loss of coherence in the
presentation.

\section{Model and qualitative description of the dynamics} 
\noindent 
We investigate theoretically the spectrum of resonance fluorescence of a
driven three--level atom, whose center--of--mass motion is confined by a
harmonic potential, as depicted in Fig.~1. For simplicity, we consider
one--dimensional motion along the $x$-axis.
The relevant electronic transitions are arranged in a
$\Lambda$--configuration, composed of two stable or metastable states $\ket 1$
and $\ket 2$ and an excited state $\ket 3$. The transitions $\ket j \to \ket
3$ are dipoles with moments ${\bf d}_j$  and linewidth $\gamma_j$ $(j=1,2)$,
such that the linewidth of the excited state $\ket 3$ is $\gamma =
\gamma_1+\gamma_2$. The atom is driven by the bichromatic field ${\bf E}={\bf
  E}_1+{\bf E}_2$, with
\begin{equation} 
{\bf E}_j(x, t)={\cal E}_{j}\epsv_j  {\rm e}^{{\rm i}k_j\cos\phi_j x} 
{\rm e}^{-{\rm i}\omega_{{\rm L},j}t}+ {\rm c.c.}~,
\label{eq:EFields}
\end{equation} 
where $x$ denotes the atomic center--of--mass position.
Here ${\cal E}_j$ and $\epsv_j$ are amplitude and polarization of the field
modes at the optical frequency $\omega_{{\rm L},j}$ with wave vector $k_j$, 
and $\phi_j$ is the angle between the laser wave vector and the
{\it x}-axis. The components ${\bf E}_j$ drive the dipoles ${\bf d}_j$ and are
tuned by the same detuning $\delta$ from resonance, such that the states $\ket
1$ and $\ket 2$ are resonantly coupled by two photon processes.  
A detector monitors the light scattered at the angle $\psi$ with respect to
the $x$-axis, thereby measuring the spectrum of the intensity. 

Throughout this work, we investigate the manifestation of the mechanical
effects of the interaction between light and atom in the spectral signal. The
system is in the regime where the size of the center--of--mass wave packet
$\Delta x$ is much smaller than the wavelength of the incident radiation
$\lambda_L$(Lamb--Dicke regime), and the laser field ${\bf E}$ cools the
motion~\cite{EITcooling,Giovanna03}.

\begin{figure} 
\includegraphics[width=6cm]{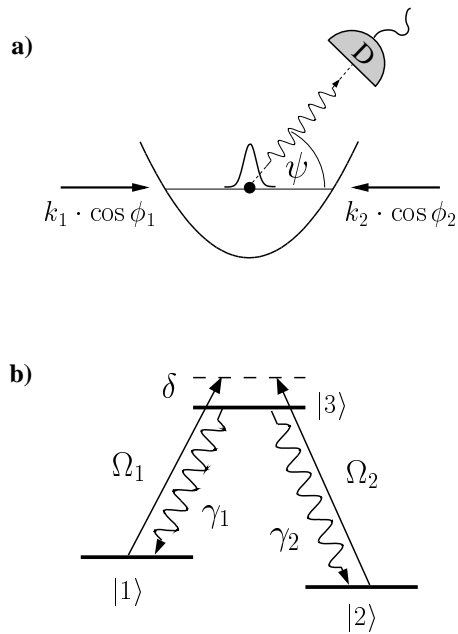} 
\caption{{\bf a)} Geometry of the lasers and position of the detector with
  respect to the axis $x$ of the atomic center--of--mass motion. We denote
  with $k_1\cos\phi_1$ and $k_2\cos\phi_2$ the projections of the laser wave
  vectors on the 
$x$-axis. The detector D records the light scattered at an angle $\psi$ with
respect to the motional axis. 
{\bf b)} Relevant electronic transitions. The stable or metastable states
$\ket{1}$ and $\ket{2}$ are coupled by dipole transitions to the excited state
$\ket{3}$. The lasers drive the transition $\ket{j}\to\ket{3}$ with Rabi
frequency $\Omega_j$  and both are tuned from resonance by $\delta$. The state
$\ket 3$ decays with rate $\gamma_j$ into $\ket{j}$ ($j=1,2$).}
\label{Fig:1}  
\end{figure}

In the Lamb--Dicke regime, the dynamics of the driven atom can be described by
a hierarchy of processes at the different orders in the ratio 
$\Delta x/\lambda_L$, which accounts for the effects of the field spatial
gradient over the center--of--mass wave packet.
At zero order in $\Delta x/\lambda_L$, internal and external degrees of freedom
are decoupled, and the internal stationary state of the atom is the
dark state~\cite{CPT}
\begin{equation}
|\psi_{\rm D}\rangle
=\frac{\Omega_2|1\rangle-\Omega_1|2\rangle}{\sqrt{\Omega_1^2+\Omega_2^2}},
\end{equation}
with the Rabi frequency ${\Omega_j={\bf d}_j\cdot \epsv_j{\cal E}_j/\hbar}$, 
which we assume to be real. In this limit, at steady state the density matrix
of the atom is the product $\rho_{\rm D}\mu$ 
of the density matrix for the external 
degrees of freedom $\mu$ and for the internal degrees of freedom
$\rho_{\rm D}=|\psi_{\rm D}\rangle \langle \psi_{\rm D}|$. 

At first order in $\Delta x/\lambda_L$ the state $|\psi_{\rm D}\rangle$
becomes unstable due to the spatial gradient of the field over the 
finite size of the wave packet: Therefore, 
at steady state the density matrix of the atom is $\rho_{\rm st}=\rho_{\rm
  D}\mu+{\rm O}(\Delta x/\lambda_L)$, where the correction ${\rm O}(\Delta
x/\lambda_L)$ accounts for the processes due to the mechanical effects of the
coupling between light and atom. 

The dynamics of this system has
been investigated in~\cite{Giovanna03} in the context of laser-cooling. There,
it has been characterized by two main time scales: A faster time
scale $T_0$, for the scattering processes  
at zero order in $\Delta x/\lambda_L$, where internal and external degrees of
freedom are decoupled, and a slower time scale $T_{1}\gg T_0$, where the
effects due to the field gradient along the center--of--mass wave packet
manifest. On the time scale $T_0$ the internal dynamics accesses the dark
state $\ket{\psi_D}$, and the atom ceases to scatter photons. On the time
scale $T_{1}$, the atom absorbs light that is out
of phase with the laser field due to the harmonic motion, 
thereby leaving the dark state and undergoing
transitions that change the vibrational state. Then, light is scattered at
zero order in $\Delta x/\lambda_L$ and the atom reaccesses the dark state. In
other words, on a coarse--grained dynamics the internal state of the atom is
$\rho_{\rm D}$, whereas the correction term ${\rm O}(\Delta x/\lambda_L)$ in
$\rho_{\rm st}$ accounts for the processes which couple internal and external
degrees of freedom, and give rise to the spectrum of emission.  

In Fig.~2 two spectra of emission of the dipole ${\bf d_1}$~\cite{Footnote} 
are shown for (a)
$\Delta x/\lambda_L\approx 7 \times 10^{-4}$  
and (b) $\Delta x/\lambda_L\approx 4 \times 10^{-3}$. The most striking feature
is the visibility of the sidebands of the elastic peak, namely the two
signals at $\omega_{L,1}\pm \nu$, compared to the rest of the spectrum. These
motional sidebands are Lorentz curves of equal height, and their functional
dependence on the frequency is plotted in the insets of Fig.\ 2. They
originate from Raman scattering processes, where the initial and final state is
$|\psi_{\rm D}\rangle$ and the vibrational state is changed by one
phonon. These processes occur on the time scale $T_{1}$, which also determines
the linewidth of the resonances.  

The broad signals of the spectrum correspond to the Mollow---type inelastic
spectrum~\cite{Narducci90,Plenio95} 
and their typical height is orders of magnitude smaller than the
height of the motional sidebands for $\Delta x/\lambda_L\approx 7\times
10^{-4}$, while it is 
comparable for $\Delta x/\lambda\approx 4\times 10^{-3}$. We remark that the
intensity of the elastic peak is at higher order in $\Delta x/\lambda_L$. In
the following, we evaluate and discuss the spectrum in detail.

\begin{figure}[h]
\includegraphics[width=7cm]{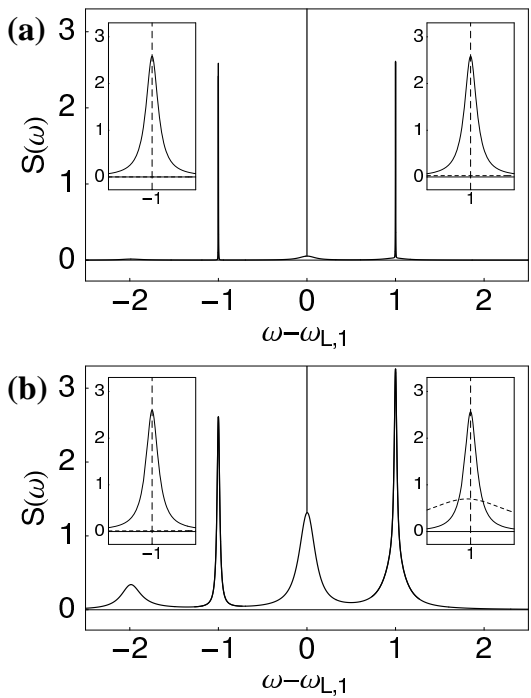}
\caption{Spectrum of resonance fluorescence $S(\omega)$ (in arbitrary units)
as a function of the frequency $\omega-\omega_{L,1}$ in units of $\nu$. The parameters are  $\Omega_1=\Omega_2=8.5\nu$, $\gamma=10\nu$, $\gamma_1=\gamma_2$, $\delta=35\nu$, $\phi_1=0$, $\phi_2=\pi$, corresponding to $\langle
  n\rangle=0.005$.  In 
{\bf a)} $\eta_1=\eta_2=0.01$, in {\bf b)} $\eta_1=\eta_2=0.05$. In both
figures, the insets show the details of the curves at $\omega=\omega_{L,1}\pm \nu$.
The contributions to the spectrum are plotted separately: The solid line
  corresponds to the signal of the Stokes and anti-Stokes components, the
  dashed line to the signal due to the Mollow--type inelastic spectrum.} 
\label{Fig:2}
\end{figure}

\section{Evaluation of the spectrum of resonance fluorescence}

\noindent 
Let the detector measure the radiation scattered by the dipole ${\bf d_1}$
\cite{Footnote}. In the far--field the spectrum at frequency $\omega$ is
determined by ${\cal S}(\omega) = \chi S(\omega)$, where $\chi$ collects all
prefactors which do not depend on $\omega$. All results are rescaled by this
common factor, which contains the dipole radiation pattern and is therefore a
function of $\psi$. At a different angle than the laser propagation
directions, the expression 
\begin{equation} 
\label{SomegaPol:2} 
S(\omega)= {\rm Re}\int_0^{\infty}{\rm d}\tau{\rm e}^{-{\rm i}(\omega-\omega_{L,1})\tau} \langle D^\dagger(x,\tau) D(x,0)\rangle
\end{equation} 
contains the frequency dependence, where $D(x,t)$ is the generalized dipole
lowering operator for the transition $\ket 1 \to \ket 3$ in the reference frame
rotating with the laser frequency $\omega_{L,1}$. By means of the quantum
regression theorem, the two--time correlation function in
Eq.~(\ref{SomegaPol:2}) is determined by 
the Liouvillian ${\cal L}$ defined in the master equation
$\partial\rho/\partial t={\cal L}\rho$ for the atomic
dynamics~\cite{Carmichael93}. Thus, the
dipole lowering operator at time $t$ in Eq.~(\ref{SomegaPol:2}) 
is given by $D(x,t)=D(x) {\rm e}^{{\cal
    L} t}$ with $D(x)={\rm e}^{-{\rm i}k_1 x \cos\psi}|1\rangle\langle
3|$. Here, the center-of-mass position $x$ is an operator acting on the atomic
external degrees of freedom. The average $\langle \cdot \rangle$ is taken over
the atomic density matrix $\rho_{\rm st}$ at steady state, given by ${\cal
  L}\rho_{\rm st}=0$. 

We evaluate $S(\omega)$, Eq.~(\ref{SomegaPol:2}), by applying the spectral decomposition of the Liouville operator ${\cal L}$~\cite{Briegel93,Barnett2000} according to the secular equations
\begin{eqnarray*}
{\cal L}\rho^{\lambda}&=&\lambda\rho^{\lambda},\\
\check{\rho}^{\lambda}{\cal L}&=&\lambda \check{\rho}^{\lambda},
\end{eqnarray*}
with eigenvalues $\lambda$ and right and left eigenelements $\rho^{\lambda}$
and $\check{\rho}^{\lambda}$, respectively. The orthogonality and completeness
of the eigenelements is defined with respect to the trace, such that ${\rm
  Tr}\{\check{\rho}^{\lambda^{\prime}}\rho^{\lambda}\} = 
\delta_{\lambda^{\prime},\lambda}$, where any density operator $\rho$ can be 
decomposed as $\rho=\sum_{\lambda}\rho^{\lambda}{\rm Tr}
\{\check{\rho}^{\lambda}\rho\}$~\cite{FootnoteCompleteness}.
We define the projectors onto the eigenspace corresponding to the eigenvalue
$\lambda$ as ${\cal P}^{\lambda}=\rho^{\lambda}\otimes \check\rho^{\lambda}$
leading to 
\begin{equation} 
{\cal L}{\cal P}^{\lambda}={\cal P}^{\lambda}{\cal L}=\lambda{\cal
  P}^{\lambda}.  
\end{equation} 
Their action on an operator $X$ is defined as ${\cal
  P}^{\lambda}X=\rho^{\lambda}{\rm Tr}\{\check{\rho}^{\lambda}X\}$. By
applying this formalism, we rewrite Eq.~(\ref{SomegaPol:2}) as  
\begin{equation} 
\label{S:3} S(\omega)={\rm Re} \sum_{\lambda}\frac{1}{{\rm
i}(\omega-\omega_{L,1})-\lambda} {\rm  Tr}\left\{D^{\dagger}(x)
{\cal P}^{\lambda}D(x)\rho_{\rm st}\right\}.
\end{equation}  
Hence, the spectrum of resonance fluorescence is the sum of Lorentz and/or
Fano--like 
profiles, centered at the imaginary part of the eigenvalues of ${\cal L}$, with
width given by the real part of $\lambda$. The exact evaluation of the
spectrum for this kind of problem, with an infinite number of degrees of
freedom, is a hard task. Nevertheless, an analytic solution can be found in
the Lamb--Dicke regime. In this limit, we make
perturbation theory in the parameter $\Delta x/\lambda_L$ 
on a spectral decomposition of ${\cal L}$, of which the
spectrum $\{\lambda\}$ and the respective eigenelements at zero order are
known.  

\subsection{Theoretical description} 
\noindent
The master equation for the atomic dynamics reads
\begin{equation}
\label{master} 
\frac{\partial}{\partial t}\rho(t)={\cal L}\rho(t)=\frac{1}{{\rm i}\hbar}[H,\rho(t)]+{\cal K}\rho(t),
\end{equation}
where $H$ is the Hamilton operator for the coherent dynamics and ${\cal K}$ is
the Liouvillian describing spontaneous emission. We decompose the Hamilton
operator as 
\begin{eqnarray*}
H=H_{\rm  mec}+H_{0}+V(x),
\end{eqnarray*}
where $H_0=\hbar\delta\sum_{j=1,2}|j\rangle\langle j|$ gives the eigenenergies
of the electronic states in the reference frame of the laser, with
$\delta=\omega_{L,1}-\omega_1=\omega_{L,2}-\omega_2$, where $\omega_j$ denotes
the frequency of the transition $\ket j \to \ket 3$. The term $H_{\rm mec}$
describes the center of mass motion of the atom with mass $M$ in a harmonic
potential of frequency $\nu$, 
\begin{equation} 
H_{\rm  mec}=\frac{p^2}{2M}+\frac{1}{2}M\nu^2x^2= \hbar\nu\left(a^{\dagger}a+\frac{1}{2}\right),
\end{equation}  
where $x$ and $p$ are the canonical conjugate variables describing position
and momentum of the atom, whereas $a$ and $a^{\dagger}$ are the annihilation,
creation operators of a quantum of energy $\hbar\nu$, respectively, such that
$x=\sqrt{\hbar/2M\nu}(a+a^{\dagger})$, and $p={\rm i}\sqrt{\hbar
  M\nu/2}(a^{\dagger}-a)$. We denote with $|n\rangle$ the eigenelements of
$H_{\rm mec}$, fulfilling 
$H_{\rm mec}|n\rangle=\hbar \nu (n+1/2)|n\rangle$ with $n=0,1,2,\ldots$ 

The term $V(x)$ describes the coherent interaction of the atom with the lasers at the position $x$ of the center of mass, \begin{equation} 
\label{dipoleinteraction}
V(x)=\frac{1}{2}\sum\limits_{j=1,3}\hbar\Omega_{j} \left({\rm e}^{-{\rm i}
    k_j\cos\phi_jx}|3\rangle\langle j|+{\rm h.c.}\right), 
\end{equation} 
and the exponentials account for the recoil
momenta $\pm \hbar k_j\cos\phi_j$  of the atom when absorbing or emitting a
photon. The operator ${\cal K}$ in Eq.~(\ref{master}) describes the
spontaneous decay, according to
\begin{eqnarray} 
\label{spontaneousemission} 
{\cal K}\rho(t)=&&-\frac{\gamma}{2}
    \left(\ibrkt{3}{3}\rho(t)+\rho(t)\ibrkt{3}{3}\right)\nonumber\\
&&+\sum\limits_{j=1,2}
    \gamma_j\ibrkt{j}{3}\tilde{\rho}_j(t)\ibrkt{3}{j},
\end{eqnarray} 
where $\gamma_1+\gamma_2=\gamma$. Here, we have introduced 
\begin{equation}
\label{momentumtransfer}\tilde{\rho}_j(t)=\int\limits_{-1}^{1}{\rm
    d}\!\cos\!\theta \, {\cal N}(\cos\!\theta) \, 
{\rm e}^{ik_jx\cos\theta}\rho(t)\;{\rm
    e}^{-ik_jx\cos\theta},
\end{equation} 
which describes the momentum transfer $\hbar k_j\cos\theta$
    due to the photons spontaneously emitted at angle $\theta$ with respect
to the motional axis and with angular distribution ${\cal N}(\cos\theta)$. \\ 

\subsection{Perturbative expansion in the Lamb--Dicke parameter}
    \label{sec:pertth} 
\noindent 
In the Lamb--Dicke regime we can approximate
$\exp\left(\pm {\rm i}k_jx\cos\varphi\right)=(1-\eta_j^2\cos^2\varphi/2(2a^{\dagger}a+1))\pm {\rm i}\eta_j
    \cos\varphi(a^{\dagger}+a)+{\rm O}(\eta_j^3)$, where 
$$\eta_j=\sqrt{\frac{\hbar
    k_j^2}{2M\nu}}$$ 
is the Lamb--Dicke parameter, corresponding to the ratio of
    the size of the oscillator ground state over the laser wavelength. Here,
    $\eta_j$ 
    is the parameter of the perturbative expansion, and it fulfills the
    relation $\eta_j\sqrt{2\langle n\rangle+1}\approx \Delta x/\lambda_L \ll 1$. In second order in $\eta_j$  
expression~(\ref{S:3}) has the form 
\begin{equation}
    S(\omega)=S_{0}(\omega)+S_{1}(\omega)+S_{2}(\omega)+{\rm O}(\eta_j^3),
\end{equation} where the subscript $\alpha=0,1,2$ indicates the corresponding
order in the perturbative expansion. In order to evaluate
$S_{\alpha}(\omega)$, we expand the operators ${\cal L}$ and $D$ in power of $\eta_j$,
yielding
\begin{eqnarray*} &&D_0=|1\rangle\langle 3|,\\ 
&& D_1=-{\rm  i}k_1x\cos\psi|1\rangle\langle 3|,\\
&&D_2=-\frac{1}{2}k_1^2x^2\cos^2\psi|1\rangle\langle 3|, 
\end{eqnarray*}
and 
\begin{eqnarray} 
{\cal L}_{0}\rho
    &=&\frac{1}{{\rm i}\hbar} [H_{\rm
    mec},\rho]+\frac{1}{{\rm i}\hbar}
[H_0+V(0),\rho]+{\cal K}_0\rho\nonumber\\ &=&({\cal L}_E+{\cal
    L}_I)\rho, \label{LI:LE}\\ {\cal L}_{1}\rho &=&\frac{1}{{\rm
    i}\hbar}\left[V_1x,\rho\right],\label{L:1}\\
%&=&-{\rm i}\sum_{j=1}^2\eta_j\Omega_j\cos\phi_j[(|3\rangle\langle j|-
%|j\rangle\langle 3|)(a+a^{\dagger},\rho]\nonumber\\
{\cal L}_{2}\rho &=&
    \frac{1}{2{\rm    i}\hbar}\left[V_{2}x^2,\rho\right]+{\cal
    K}_{2}\rho. \label{L:2}
\end{eqnarray}
In Eq.~(\ref{LI:LE}) we have introduced the Liouville operators ${\cal L}_E$
    and ${\cal L}_I$, which account for the external and internal degrees of
    freedom. They are defined as ${\cal L}_E\rho=1/{\rm i}\hbar 
    \left[H_{\rm mec},\rho\right]$ and  ${\cal L}_I\rho=1/{\rm i}\hbar
    \left[H_0+V(0),\rho\right]+{\cal K}_0\rho$. 
 
By determining ${\cal L}_\alpha$, we have used the expansion of the
interaction term, Eq.~(\ref{dipoleinteraction}), $V(x)=V(0)+V_{1}x
+V_{2}x^2/2+{\rm O}(\eta_j^3)$ with  
\begin{equation}
V_{\alpha}=\frac{\partial^{\alpha}}{\partial
  x^{\alpha}}V(x)\Bigl|_{x=0},~\alpha=1,2 
\end{equation}
and the expansion of Eq.~(\ref{spontaneousemission}), yielding  
\begin{eqnarray*} &&{\cal
    K}_{0}\rho=\sum\limits_{j=1,2}\frac{\gamma_j}{2}
\left(2\ibrkt{j}{3}\rho\ibrkt{3}{j}
    -\ibrkt{3}{3}\rho-\rho\ibrkt{3}{3}\right),\\  
&&{\cal K}_{2}\rho=\beta \sum\limits_{j=1,2}\gamma_j
    k_j^2\;\ibrkt{j}{3}\left(2  x\rho x-x^2\rho-\rho
    x^2\right)\ibrkt{3}{j} \nonumber. 
\end{eqnarray*} 
Here,  $\beta=\int_{-1}^{+1}{\rm d}\!\cos\!\theta~{\cal N}(\cos\!\theta)
\cos^2\!\theta$
is a constant. We remark that the first order term ${\cal K}_1$ vanishes after
averaging over the angles of emission $\theta$.  

From Eq.~(\ref{LI:LE}) it can be seen that internal and external degrees of freedom are decoupled at zero order. Hence, the eigenvalues of ${\cal L}_0$ are 
\begin{equation} 
\lambda_0= \lambda_E+\lambda_I,  
\end{equation} with
$\lambda_E$ and $\lambda_I$ being the eigenvalues of ${\cal L}_E$ and ${\cal L}_I$, respectively. The projector ${\cal P}^{\lambda}_0$ in the corresponding eigenspace factorizes into the projectors ${\cal P}^{\lambda_I}_I$ and  ${\cal P}^{\lambda_E}_E$ assigned to the internal and external degrees of freedom, according to
\begin{equation} \label{PI:PE}
{\cal  P}^{\lambda}_0
= {\cal P}^{\lambda_I}_I{\cal P}^{\lambda_E}_E.
\end{equation}
The spectrum of ${\cal L}_I$ characterizes the 
dynamics of the three-level transition and the spectral properties of
the radiation emitted by the bare atom. 
The eigenvalues of ${\cal L}_E$ take on the values 
$\lambda_E={\rm i}\ell\nu$, with $\ell=0,\pm 1,\pm 2,\ldots$ Each eigenspace
at $\lambda_E$
is infinitely degenerate, and 
the corresponding left and right eigenelements
are, for instance,
$\check{\mu}_n^{\ell}=|n+\ell\rangle\langle n|$, $\mu^{\ell}_n=|n\rangle\langle
    n+\ell|$. These eigenelements constitute a complete and orthonormal
    basis over the eigenspace at this eigenvalue. In particular, the projector over the eigenspace at $\lambda_E={\rm i}\ell\nu$ is defined on an operator $X$ as
\begin{eqnarray}
\label{PE}
{\cal P}^{\lambda_E={\rm i}\ell\nu}_EX
&=&\sum_n \mu^{\ell}_n {\rm Tr}_E
\{\check{\mu}_n^{\ell}X\}\\
&=&\sum_n |n\rangle\langle n|X|n+\ell\rangle\langle n+\ell|,\nonumber
\end{eqnarray}
where ${\rm Tr}_E$ denotes the trace over the external degrees of freedom.

At first order in the expansion in $\eta_j$, internal and external degrees of
freedom are coupled, and the degeneracy of the subspaces at eigenvalue
$\lambda_E$ is lifted \cite{Lindberg84,Stenholm86}. 
The perturbative corrections to the eigenvalues $\lambda_0$, to the 
eigenelements $\rho^{\lambda}_0$, $\check{\rho}^{\lambda}_0$, and to 
the projectors ${\cal P}^{\lambda}_0$ are found by solving iteratively the 
secular equations at the same order $p$ in the perturbative expansion, that is 
\begin{eqnarray}
\label{Sec:Pert:1}
    &&\sum_{\alpha=0}^p {\cal
    L}_\alpha\rho^{\lambda}_{p-\alpha}=\sum_{\alpha=0}^p\lambda_\alpha\rho^{\lambda}_{p-\alpha},\\
    &&\sum_{\alpha=0}^p  \check{\rho}^{\lambda}_{p-\alpha}{\cal  L}_\alpha=\sum_{\alpha=0}^p
    \lambda_\alpha\check{\rho}^{\lambda}_{p-\alpha},
\label{Sec:Pert:2}
\end{eqnarray} 
where $\rho_\alpha^\lambda$ and $\check\rho_\alpha^\lambda$ are the $\alpha$--order corrections to the eigenelements $\rho_0^\lambda$ and $\check\rho_0^\lambda$.
The explicit forms up to second order are 
derived in Appendix A. In particular, $\rho_0^{\lambda=0}=\rho_{\rm D}\mu$ is the steady-state density matrix
at zero order, where $\mu$ is the density matrix for the external degrees of
freedom in the final stage of the laser--cooling dynamics~\cite{Footnote:mu}, 
and has the form
\begin{equation} 
\label{mu}
\mu=\frac{1}{1+\langle
    n\rangle}\left( \frac{\langle n\rangle}{1+\langle
      n\rangle}\right)^{a^{\dagger}a},
\end{equation}
where  
\begin{equation}
\langle n\rangle={\rm Tr}\{a^{\dagger}a\mu\}
\end{equation}   
is the average phonon number at steady state.

By substituting the explicit form of the operators into Eq.~(\ref{S:3}), we find that the zero-- and first--order contributions to the spectrum vanish (see discussion in Appendix B), yielding \begin{equation}
    S(\omega)=S_2(\omega)+{\rm O}(\eta_j^3), \label{eq:ExpS}
\end{equation} 
with
\begin{equation}
S_2(\omega)=\sum_{\lambda}\frac{g(\lambda)}{{\rm
    i}(\omega-\omega_{L,1})-\lambda}. \label{S:4} 
\end{equation} 
Here, $g(\lambda)$ is a complex-valued function, which we decompose for
convenience into $g(\lambda)=f^{(1)}(\lambda)+f^{(2)}(\lambda)$, with
\begin{eqnarray*}  
&&f^{(1)}(\lambda)={\rm
    Tr}\{D_0^{\dagger}{\cal P}_1^{\lambda}D_0
    \rho_1\},\\
&&f^{(2)}(\lambda)={\rm Tr}\{D_0^{\dagger}{\cal
    P}_0^{\lambda}D_0\rho_2\}.\end{eqnarray*} 
Using $\rho_1^{\lambda}$, $\rho_2^{\lambda}$ and ${\cal P}_1^{\lambda}$,
evaluated in Appendix A, and making use of 
    relation~(\ref{PI:PE}), we separate the
    trace terms $f^{(1)}$ and $f^{(2)}$ into the product of the trace over the external and over the internal (${\rm Tr}_I\{\}$) degrees of freedom.
By applying the cyclic properties of the trace and the the completeness relation for the external degrees of freedom, $\sum_{\lambda_E}P_E^{\lambda_E}=\uop_E$, we find 
\begin{eqnarray*}
    &&f^{(1)}(\lambda)=\frac{1}{\hbar^2}\sum_{\lambda_E'}
    \Bigl[\delta_{\lambda_E,0}{\rm   Tr}_E\left\{({\cal
    P}_E^{\lambda_E'}[x,\mu])x\right\}\\&&\times{\rm
    Tr}_I\left\{D_0^\dagger{\cal
    P}^{\lambda_I}_I\left[V_1,\frac{1}{\lambda_0-\lambda_E'-{\cal
    L}_I}D_0\frac{1}{\lambda_E' + {\cal L}_I}V_1\rho_{\rm D}\right]\right\}\\
    &&+\delta_{\lambda_E,0}{\rm Tr}_E\left\{({\cal P}_E^{\lambda_E'}\mu
    x)x\right\} \\&&\times {\rm Tr}_I\left\{D_0^\dagger{\cal
    P}^{\lambda_I}_I\left[V_1,\frac{1}{\lambda_0-\lambda_E'-{\cal
    L}_I}D_0\frac{1}{\lambda_E' + {\cal L}_I}[V_1,\rho_{\rm
    D}]\right]\right\}\\&&+\delta_{\lambda_E',0}{\rm Tr}_E\left\{({\cal
    P}_E^{\lambda_E}[x,\mu])x\right\} \\&&\times{\rm
    Tr}_I\left\{D_0^\dagger\frac{1}{\lambda_0-{\cal L}_I}\left[V_1,{\cal
    P}^{\lambda_I}_ID_0\frac{1}{\lambda_E+{\cal L}_I} V_1\rho_{\rm
    D}\right]\right\}\\ &&+\delta_{\lambda_E',0}{\rm Tr}_E\left\{({\cal
    P}_E^{\lambda_E}\mu x)x\right\} \\ &&\times {\rm
    Tr}_I\left\{D_0^\dagger\frac{1}{\lambda_0-{\cal L}_I}\left[V_1,{\cal
    P}^{\lambda_I}_I D_0\frac{1}{\lambda_E + {\cal L}_I}[V_1,\rho_{\rm
    D}]\right]\right\}\Big], 
\end{eqnarray*}  
where we have used the relation ${\rm Tr}_E\{P^{\lambda_E}_EX\}=\delta_{\lambda_E,0}{\rm Tr}_E\{X\}$.
Analogously, 
\begin{eqnarray*}&&
    f^{(2)}(\lambda)=-\frac{1}{\hbar^2}\delta_{\lambda_E,0}\sum_{\lambda_E'}\Bigl[{\rm
    Tr}_E\{({\cal P}_E^{\lambda_E'}\mu x)x\}\\&&\times{\rm
    Tr}_I\left\{D_0^\dagger{\cal P}_I^{\lambda_I}D_0{\cal
    L}_I^{-1}\left[V_1,\left(\frac{1}{\lambda_E'+{\cal L}_I}[V_1,\rho_{\rm
    D}]\right)\right]\right\}\nonumber\\ &&+{\rm Tr}_E\{({\cal
    P}_E^{\lambda_E'}[x,\mu])x\}\\ &&\times {\rm Tr}_I\left\{D_0^\dagger{\cal
    P}_I^{\lambda_I}D_0{\cal L}_I^{-1}\left[V_1,\left(\frac{1}{\lambda_E'+{\cal
    L}_I}V_1\rho_{\rm D}\right)\right]\right\}\Big]\nonumber\\
    &&-\frac{\rm i}{2\hbar}\delta_{{\lambda_E},0}{\rm Tr}_E\{\mu x^2\} {\rm
    Tr}_I\left\{D_0^\dagger {\cal P}_I^{\lambda_I} D_0 {\cal
    L}_I^{-1}[V_2,\rho_D]\right\}\nonumber. \end{eqnarray*}
From the properties of the terms ${\rm Tr}_E\{X\}$ one can already see that the eigenelements and eigenvalues of ${\cal L}_E$ contributing to the spectrum at second order are at $\lambda_E=0,\pm {\rm i}\nu$. 

In summary, the first non-vanishing contribution to 
the spectrum of emission is in second order in the perturbative
expansion. It can be decomposed into the sum of curves, centered at
the imaginary part of the eigenvalues $\lambda$ of the Liouville operator ${\cal L}$, and weighted by the factor $g(\lambda)$. The
eigenvalues are here determined up to the second order correction,
$\lambda=\lambda_0+\lambda_2+{\rm O}(\eta_j^3)$, whereby $\lambda_1=0$, as shown
in Appendix A and in~\cite{Lindberg84}. At zero order in the perturbative
expansion $\lambda_0=\lambda_I+\lambda_E$, where $\lambda_I$ is the
eigenvalue of the Liouvillian of a bare $\Lambda$--transition, while the only relevant
external eigenvalues are $\lambda_E=0,\pm {\rm i}\nu$. 

Below, we analyze the spectrum in detail. For later convenience, we rewrite $S(\omega)=S_{\rm el}(\omega)+S_{\rm M}(\omega)+S_{\rm SB}(\omega)$, where
$S_{\rm el}(\omega)$ is the contribution of the elastic peak, at $\lambda=0$, 
the term
\begin{equation}
S_{\rm M}(\omega)=\sum\limits_{\lambda_I\neq 0, \lambda_E} \frac{g(\lambda)}{{\rm i}(\omega-\omega_{L,1})-\lambda}
\label{eq:Mollow} 
\end{equation}
denotes the contributions at $\lambda_I\neq 0$, which we refer to as the
Mollow--type inelastic component~\cite{Narducci90,Plenio95}. The term
\begin{equation}        
S_{\rm SB}(\omega)=\sum\limits_{\lambda_I = 0,\atop \lambda_E=\pm{\rm i}\nu}
\frac{g(\lambda)}{{\rm i}(\omega-\omega_{L,1})-\lambda}  
\end{equation}
represents the contributions at $\lambda_I=0, \lambda_E=\pm {\rm i} \nu$, that
we identify with the sidebands of the elastic peak, or Stokes-, anti-Stokes
components of the scattered radiation. 

We remark that $S_2(\omega)$ does not depend on the position of the detector:
the terms containing the perturbative corrections
$D_1$, $D_2$ and their adjoints do not contribute to $S(\omega)$ in second
order in $\eta_j$. Furthermore, the Lamb--Dicke parameters $\eta_1$, $\eta_2$
appear always in the form $\eta_1\cos\phi_1$, $\eta_2\cos\phi_2$, since ${\cal 
  K}_2\rho_0^{\lambda=0}\propto {\cal K}_2\rho_{\rm D}=0$ (see Appendix B). 
Therefore, the mechanical effects in second order are solely determined by the
scattering of laser photons and not by recoils due to spontaneously emitted
photons. This behaviour is due to destructive quantum interference at zero
order in the Lamb--Dicke expansion, implying that light absorption is a
first-order process~\cite{Giovanna03,FootnoteN}.

\subsubsection{The Mollow--type inelastic spectrum} 

\noindent 
At zero order in the Lamb--Dicke parameter, the eigenvalues $\lambda_I\neq 0$
determine the position and the shape of the contributions to the Mollow--type
inelastic spectrum. As $\lambda$ appears in the denominator of
$S_{\rm M}(\omega)$, the second-order correction $\lambda_2$ can be neglected. 

The trace terms over the external degrees of freedom are
conveniently evaluated using the basis set corresponding to the projectors in
Eq.\ (\ref{PE}), giving
\begin{eqnarray}
\nonumber
&&{\rm
    Tr}_E\{({\cal P}_E^{\lambda_E} \mu x)x\}
=x_0^2\left[\delta_{{\lambda_E},{\rm i}\nu}(\langle n
    \rangle+1)+\delta_{{\lambda_E},-{\rm i}\nu}\langle n  \rangle\right],\\
\nonumber
&&{\rm
    Tr}_E\{({\cal P}_E^{\lambda_E} [x,\mu])x\}=
    x_0^2\left[-\delta_{{\lambda_E},{\rm i}\nu}
+\delta_{{\lambda_E},-{\rm i}\nu}\right],\\  
\label{PE:x}
&&{\rm Tr}_E\{x^2\mu\} = x_0^2\left[2 \langle n  \rangle +1\right],
\end{eqnarray} 
with $x_0=\sqrt{\hbar/2 M\nu}$. 
Thus, at second order in the Lamb--Dicke expansion the eigenelements of ${\cal
  L}_E$ determining the spectrum are in the eigenspaces corresponding to $\lambda_E=0,\pm
{\rm i}\nu$. 
Since $S_{\rm M}(\omega)$ is linear in these terms, this component of the
spectrum scales with $\eta_j^2\cos^2\phi_j$ and is linear in the average
phonon number $\langle n\rangle$. 

This spectral component is constituted by the contributions of the signals centered at ${\rm Im}\{\lambda_I\}$ and at ${\rm Im}\{\lambda_I'\}\pm\nu$. The latter originate from the term $f^{(1)}(\lambda)$ in $g(\lambda)$. 
We illustrate this behaviour in 
Fig.\ 3, where the Mollow--type inelastic spectrum is shown for different values of the detuning $\delta$ (and correspondingly of the average phonon number at steady
state $\langle n\rangle$). The frequencies ${\rm Im}\{\lambda_I\}$  are marked with crosses on the frequency axis, where the arrows indicate the center--frequencies ${\rm Im}\{\lambda_I'\}$ of which only the sidebands are visible. 
In order to highlight this splitting, we have taken borderline parameters,
such that all peaks are clearly resolved. 
For realistic parameters, the sidebands on the left side of the spectrum are
visible, as shown in Fig.~2(b).

The curves at ${\rm Im}\{\lambda_I\}$ can be reproduced by evaluating the
spectrum of emission of a bare three-level atom, whose ground state coherence
has a finite 
lifetime~\cite{Narducci90,Plenio95}. Thus, they can be
identified with the spectrum of the photons scattered at zero order in the
Lamb--Dicke parameter. The signals centered at ${\rm Im}\{\lambda_I'\}\pm\nu$
are peculiar. They correspond to processes where photon scattering is
accompanied by a change in the vibrational state. Looking at the corresponding
eigenelements, it follows that they stem from the inelastic processes which
take the atom out of the dark state. We remark that in Fig.~2 the left pole 
${\rm Im}\{\lambda_I'\}$ falls at the same frequency as the left motional
 sideband: In fact, the parameters have been here chosen, so that the
 frequency of absorption along the cooling transition coincides with the
narrow resonance characterizing the excitation
spectrum~\cite{EITcooling,Giovanna03,Lounis92}. 
%Hence, the sidebands of this pole appearing in
%Fig.~2(b) are the sidebands of the resonance characterizing the excitation
%spectrum of a bare $\Lambda$--transition~\cite{Giovanna03,Lounis92}. 

\subsubsection{The sidebands of the elastic peak} 

\noindent 
The spectral contributions at $\lambda_I=0$, $\lambda_E=\pm {\rm i}\nu$, allow
for a compact analytic form. Only the term $f^{(1)}(\lambda)$ contributes to
$g(\lambda)$ in $S_{\rm SB}(\omega)$. After some algebraic manipulation, we write 
\begin{eqnarray*}  &&g(\lambda_E)={\rm
    Tr}_E\left\{({\cal
    P}_E^{\lambda_E}
(a+a^{\dagger})\mu)(a+a^{\dagger})\right\}|f(\lambda_E)|^2,
\label{f1:sideband}
\end{eqnarray*}  
where 
\begin{equation} 
\label{f:E}
f(\lambda_E)=\frac{x_0}{\hbar}{\rm
    Tr}_I\{D_0^{\dagger}\frac{1}{\lambda_E-{\cal L}_I}[V_1,\rho_D]
  \}. 
\end{equation}  
The explicit form~(\ref{f:E}) is found by applying the relation
$(\lambda-{\cal L}_I)^{-1}=\int\limits_0^\infty {\rm d}t\, {\rm
  e}^{-(\lambda-{\cal L}_I)t}$. Using the quantum regression theorem we arrive at
\begin{equation} 
f(\lambda_E)=-{\rm i}\frac{2\eta
    \lambda_E\Omega_1\Omega_2^2}{\Omega^2(\Omega^2+4\lambda_E
    ({\rm i}\delta+\lambda_E+\gamma/2))},
\end{equation} 
with 
$\Omega^2=\Omega_1^2+\Omega_2^2$, and where we
have introduced $\eta=x_0(k_1\cos\phi_1-k_2\cos\phi_2)$. 
The explicit
form of $S_{\rm SB}(\omega)$ is determined after evaluating the second order
corrections $\lambda_2$ to $\lambda_0=\pm {\rm  i}\nu$. 
These are found by solving the eigenvalue equations (\ref{Eq:rho:1}) and
(\ref{Eq:rho:2}) at $\lambda_0=\pm {\rm i}\nu$. In the subspace at
$\lambda_E={\rm i}\ell\nu$, $\lambda_I=0$, after tracing over the 
internal degrees of freedom, they read
\begin{eqnarray} \label{Line}\lambda_2\tilde{\mu}^{\lambda_2}&=&
  s(\nu)\left[a\tilde{\mu}^{\lambda_2} a^{\dagger}-a^{\dagger}a\tilde{\mu}^{\lambda_2}\right]
  \nonumber\\ &+& s(-\nu)\left[a^{\dagger}\tilde{\mu}^{\lambda_2}
    a-aa^{\dagger}\tilde{\mu}^{\lambda_2}\right] +{\rm H.c.}, 
\end{eqnarray} where
$\tilde{\mu}^{\lambda_2}$ are the right
eigenelements of Eq.~(\ref{Line}) at the eigenvalue $\lambda_2$.
The left eigenelements $\check{\tilde{\mu}}^{\lambda_2}$ fulfill the
corresponding
equation for the action to the left~\cite{Briegel93,Barnett2000}.
The coefficient $s(\nu)$ is given by
\begin{eqnarray}
  s(\nu)&=&\frac{1}{2M\nu}\int\limits_0^\infty {\rm d}t\, 
{\rm e}^{\iu \nu t} {\rm
    Tr}_I\{V_1{\rm e}^{{\cal L}_It} V_1\rho_{D}\}\nonumber\\&=&\eta^2\frac{{\rm
      i} \nu\Omega_{1}^2\Omega_{2}^2} {\Omega^2 (\Omega^2+4\nu
    ({\rm i}\gamma/2-\nu+\delta))}.  \end{eqnarray} 
We rewrite $s(\pm\nu)={\rm
  Re}[s(\pm\nu)]+\iu~{\rm Im}[s(\pm\nu)]$, and define $A_{\pm}=2{\rm
  Re}\{s(\mp\nu)\}$. Substituting into Eq.\ (\ref{Line}),
we obtain the more familiar form~\cite{Stenholm86,Briegel93}
\begin{eqnarray} \label{Line:Damp} \lambda_2\tilde{\mu}^{\lambda_2} 
&=&-i \bar\nu
  [a^{\dagger}a,\tilde{\mu}^{\lambda_2}]\\ 
\nonumber
&&+A_-\left[2a\tilde{\mu}^{\lambda_2}
    a^{\dagger}-a^{\dagger}a\tilde{\mu}^{\lambda_2}-\tilde{\mu}^{\lambda_2}
a^{\dagger}a\right]\\
\nonumber  
&&+A_+\left[2a^{\dagger}\tilde{\mu}^{\lambda_2}
  a-aa^{\dagger}\tilde{\mu}^{\lambda_2} 
   -\tilde{\mu}^{\lambda_2} aa^{\dagger}\right],  \end{eqnarray} 
with $\bar\nu={\rm Im}[S(\nu)]+{\rm Im}[S(-\nu)]$. 
The corresponding eigenvalues are solutions of Eq.~(\ref{lambda:2}), and have the
  form \cite{Briegel93,Stenholm86,Lindberg84}
\begin{equation} \lambda_2(N,\ell)=-{\rm
    i}\ell\bar\nu-\left(2N+|\ell|\right)(A_--A_+), 
\end{equation} 
where the index $N=0,1,2,\ldots$ accounts for the removed degeneracy inside
the eigenspace. The explicit form of the corresponding left and right
eigenelements $\check{\tilde{\mu}}^{N,\ell}$, $\tilde{\mu}^{N,\ell}$ can be 
found in \cite{Briegel93}. They form a complete and orthogonal set with
respect to the trace over the external degrees of freedom, 
${\rm
  Tr}_E\{\check{\tilde{\mu}}^{N,\ell}\tilde{\mu}^{N',\ell'}\}=\delta_{N,{N'}} \delta_{\ell,{\ell'}}$ and $\sum_{N,\ell} \tilde{\mu}^{N,\ell}\otimes
  \check{\tilde{\mu}}^{N,\ell}=\uop_E$. In particular, the eigenelements
  $\tilde{\mu}^{N,\ell}$, $\check{\tilde{\mu}}^{N,\ell}$ form a complete basis
  over the subspace at eigenvalue $\lambda_E={\rm i}\ell\nu$, such that  
$${\cal P}_E^{\lambda_E={\rm i}\ell\nu}=\sum_N \tilde{\mu}^{N,\ell}
\otimes \check{\tilde{\mu}}^{N,\ell}.$$ 
We remark that the density operator given in Eq.~(\ref{mu}) is right
eigenelement at $N=\ell=0$, that is $\mu=\tilde{\mu}^{0,0}$.  Using this basis for evaluating the trace terms over the external degrees of freedom, we get 
\begin{eqnarray} &&S_{\rm SB}(\omega) 
    ={\rm Re}
\sum_{\ell,N}\frac{|f(\lambda_E)|^2}
{{\rm i}(\omega-\omega_{L,1}-\ell\nu)-\lambda_2^{N,\ell}}\nonumber\\
    &&\times ~ {\rm
      Tr}_E\{(a+a^{\dagger})\tilde{\mu}^{N,\ell}\} {\rm
      Tr}_E\{\check{\tilde{\mu}}^{N,\ell}(a+a^{\dagger})\mu\}.  
\nonumber
\end{eqnarray} 
By using the explicit form of the eigenelements in~\cite{Briegel93}
we find that only the terms at $N=0$, $\ell=\pm 1$ contribute to the sum,
giving
\begin{equation}
\label{Stokes}
    S_{\rm SB}(\omega)=\sum_{\ell =\pm 1}\frac{\gamma_{\rm
      S}^2}{[\omega-\omega_{L,1}+\ell 
      (\nu+\bar{\nu})]^2+\gamma_{\rm S}^2}~s_0,
\end{equation} where $\gamma_{\rm
    S}=A_--A_+$ and $s_0$ is the height at the center frequency and has the 
form~\cite{footnoteOtherTransition} 
\begin{equation} s_0=\frac{\nu \Omega_1^2\Omega_2^4} {4
      \tilde{\gamma}_{\rm S}\delta \Omega^4 (\Omega^2-4 \nu^2)}. \end{equation}
Here, we have defined $\tilde{\gamma}_{\rm S}=\gamma_{\rm S}/\eta^2$, which is
at zero order in the Lamb--Dicke expansion.  

From Eq.~(\ref{Stokes}) one sees that both sidebands of the elastic peak have
the same form, independent of the angle $\psi$ of the detector with respect to
the axis of the motion. In particular,  they have the same Lorentzian shape,
as shown in the insets of Fig.~2, and are centered at the frequency $\pm(\nu+\bar{\nu})$,
where $\bar{\nu}$ is a shift in second order in the Lamb--Dicke parameter. We
illustrate this effect in Fig.\ 4, where we have chosen suitable parameter to
show this small shift most clearly. Here, $\bar \nu$ can be identified
with the a.c.--Stark shift arising from off-resonant coupling to other dipole
transitions at different vibrational numbers.   
The width $\gamma_{\rm S}$ of the sidebands is at
second order in the Lamb--Dicke parameter and corresponds to the cooling
rate~\cite{Giovanna03}.  
The height $s_0$ is in zero order in the perturbative expansion. With some
algebraic manipulations, using $\langle n\rangle=A_+/\gamma_{\rm S}$, it can
be rewritten as  
\begin{equation} s_0=\frac{\Omega_2^2}{\gamma\Omega^2}\langle
    n\rangle (1+\langle n\rangle ).  
\end{equation}
Thus, the motional sidebands are well distinguished compared to the Mollow--type
inelastic spectrum for ${\eta^2\ll \langle n\rangle}$.  

\begin{figure}
\includegraphics[width=7cm]{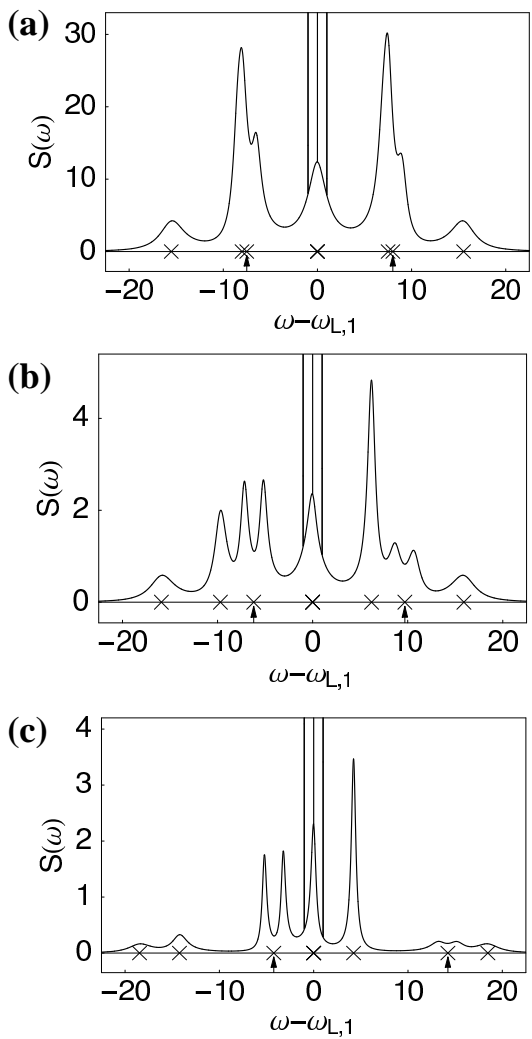}
\caption{Spectrum $S(\omega)$ in arbitrary units as a function of
  $\omega-\omega_{L,1}$ in units of $\nu$. The figures are 
at different values of the detuning $\delta$, for
$\Omega_1=\Omega_2=10\nu$, $\gamma=5\nu$, $\gamma_1=\gamma_2$, $\phi_1=0,
\phi_2=\pi$, $\eta_1=\eta_2=10^{-4}$. {\bf (a)} $\delta = 0.5\nu$, 
corresponding to $\langle n\rangle = 30$. {\bf (b)} $\delta = 3.5\nu$,
corresponding to $\langle n\rangle =3.8$. {\bf (c)} $\delta = 10\nu$,
corresponding to $\langle n\rangle = 1$. In all figures, the crosses on the
frequency axis indicate the positions of the frequencies ${\rm
  Im}\{\lambda_I\}$. The arrows indicate the frequencies ${\rm
  Im}\{\lambda_I'\}$ of
which only the sidebands appear in the spectrum.}
\label{Fig:3}
\end{figure}

\subsubsection{The elastic peak} 

\noindent 
The contribution at the eigenvalue $\lambda=0$ corresponds to the elastic peak, i.e.\ to the
coherent part of the spectrum. In this system its appearance
is due to the mechanical effects of light: In fact, at zero order in the
Lamb--Dicke expansion there is no photon emission at steady state. 
We evaluate the radiation scattered at this frequency starting from
expression~(\ref{S:3}), 
\begin{eqnarray}  &&S_{\rm el}(\omega) = \pi \delta(\omega-\omega_{L,1}) \,\Tr\left\{ D^\dagger \Pj^{\lambda=0}
    D\rho_{\rm st}\right\}\nonumber\\ 
&&=\pi \delta(\omega-\omega_{L,1}) 
\,|\Tr\left\{D^\dagger\rho_{\rm st}\right\}|^2, \end{eqnarray}
which is the well-known form of the elastic peak contribution~\cite{AtomPhoton}.
The perturbative expansion of $D$ and $\rho_{\rm st}$ can now be applied for
evaluating the average dipole moment ${\rm Tr}\{D\rho_{\rm st}\}$. 
We find the first non-vanishing contribution at
O($\eta_j^2$), such that ${\rm Tr}\{D\rho_{\rm st}\}=
\Tr\{  D_0^\dagger\rho_2
\}+\Tr\{D_1^\dagger\rho_1\}+{\rm O}(\eta_j^4)$.  
This signal is due to the lowest order corrections of the Debye--Waller factor,
$\exp(-\eta_j^2\cos^2\phi_j/2)$,  on the transitions  $|\psi_{\rm
  D},n\rangle\to |3,n \rangle\to |\psi_{\rm
  D},n\rangle$, 
and to the coefficient $\eta_j^2 n\cos\phi_j\cos\psi$
on the transitions $|\psi_{\rm D},n\rangle\to |3,n\pm 1\rangle\to |\psi_{\rm
  D},n\rangle$. Thus, the intensity of the radiation scattered at the elastic peak is at
fourth order in the Lamb--Dicke parameter and it
depends on the angle of emission. 
The finite life time of the
dark state suggests also a broadened signal at this frequency.
Our analysis shows that such contribution is of higher order in the perturbative expansion.

\section{Summary of the results and discussion} 

\noindent 
The spectrum of emission of a trapped ion, whose internal degrees of freedom
constitute a $\Lambda$--transition driven at two-photon resonance, is a
remarkable
manifestation of the mechanical effects of light: In fact, at steady state
photons are emitted due to processes where the 
vibrational degrees of freedom are
excited by absorption of a photon. We have evaluated the spectrum using
perturbation theory 
in the Lamb--Dicke parameter. According to our results, the signal of the
emission spectrum is in second order in the Lamb--Dicke expansion.
We classify the
spectral features into three main contributions, which we summarize
below.  

At the laser frequency
$\omega_{L,1}$
the spectrum exhibits a 
$\delta$-peaked signal, visible for instance in Fig.\ 2, 
that we identify with 
the elastic peak. This signal is at fourth order in
the perturbative expansion. This order of magnitude is understood, as the
dipole moment at steady state scales with $\eta_j^2\sim
(\Delta x/\lambda_L)^2$. In fact, this signal is
due to Rayleigh scattering processes where the initial and final state is
the dark state $|\psi_{\rm  D}\rangle$ and the vibrational number $n$ is
conserved. Thus, this excitation originates from the lowest order mechanical 
corrections to the Rabi frequency and it depends on the angle of emission $\psi$.

At the frequencies $\omega_{L,1}\pm \nu$ one observes two narrow 
resonances. These are the motional sidebands of the
elastic peak, the Stokes and anti-Stokes components of the scattered light. 
They correspond to Raman scattering where the initial and final internal state
is $\ket{\psi_{\rm D}}$ 
and the vibrational number is changed by one phonon. The
curves are Lorentz curves, whose dependence on the physical parameters is 
given by Eq.\ (\ref{Stokes}) and plotted in the insets of Fig.~2. Their
width is in second order in the Lamb--Dicke expansion and corresponds with
the cooling rate~\cite{Stenholm86}.
The height of the curves at the center frequency scales with the average
phonon number $\langle n\rangle$ according to
$\langle n\rangle(1+\langle n \rangle)$, 
so that the total intensity emitted at this
frequency is proportional to $\eta^2\langle n\rangle$. Finally, the center 
frequencies of the sidebands are shifted from the zero--order center frequency
by a 
contribution $\bar \nu$ at second order in the Lamb--Dicke expansion: This
corresponds to 
the a.c.--Stark shift of the ground--state coherence due to the off-resonant
coupling with the excited state. Figure~4 shows the shift $\bar\nu$ for one
sideband. 

\begin{figure}
\includegraphics[width=7cm]{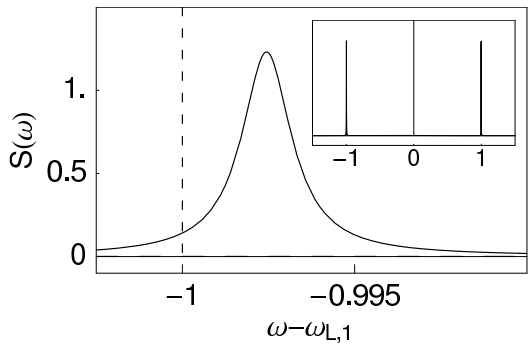}
\caption{Spectrum of resonance fluorescence $S(\omega)$ (in arbitrary units)
  as a function of the frequency $\omega-\omega_{L,1}$ in units of $\nu$. 
The parameters are $\phi_1=0$, $\phi_2=\pi$, $\eta_1=\eta_2=0.05$, 
$\Omega_1=\Omega_2=8.5\nu$, $\gamma=10\nu$, $\gamma_1=\gamma_2$,
$\delta=15\nu$, corresponding to $\langle n\rangle=0.2$. In the onset, the
Stokes component is shown: The vertical dashed line indicate the position
of the frequency $\omega-\omega_{L,1}=-\nu$. In the inset, the whole 
spectrum is shown. The signal of the 
inelastic part is here two orders of magnitude smaller than the sidebands.}
\label{Fig:4}
\end{figure}

The other spectral features, visible for instance in Fig.\ 2(b), 
can be identified with the 
Mollow--type inelastic spectrum. These can be decomposed into the sum of
Lorentz curves, whose height scales with $\eta^2$ and is linear in
$\langle n\rangle$, whereas the width is in zero order in the Lamb--Dicke expansion. 
Part of these features can be reproduced by evaluating the incoherent spectrum
of a bare $\Lambda$--transition driven at two photon resonance, whose
ground--state 
coherence has a finite decay time~\cite{Narducci90,Plenio95}. 
Hence, their origin can be explained with 
photon scattering at zero order in the Lamb--Dicke parameter, occurring once
the atom has left the dark state. Nevertheless, this part of the spectrum
exhibits also peculiar features, which cannot be understood in these
terms. These are in fact curves which characterize the excitation
spectrum of the bare $\Lambda$--atom~\cite{Lounis92}, 
and which here appear shifted by the
frequency $\pm \nu$ from their center-frequency. They thus describe scattering
processes where the vibrational number is changed by one phonon. For
saturating driving fields, they can be interpreted as Raman scattering
processes, where the initial state is the ground--state coherence $|\psi_{\rm
  D}\rangle$ and the final state is another dressed state at a different
vibrational number state. The linewidth of the emitted photon is then the
linewidth of the corresponding dressed state transition, while the center
frequency is the corresponding a.c.--Stark shift. Here, the center frequencies
are shifted by the trap
frequency $\nu$, since the dark state is excited only by processes changing the vibrational number.

Remarkably, in second order in the Lamb--Dicke expansion the spectrum $S(\omega)$ does not depend on the position of the detector, apart 
for the dipole pattern of emission. This is another consequence
of the fact that at zero order in the perturbative expansion the atom at
steady state is decoupled from radiation because of quantum
interference~\cite{FootnoteN}.

Using these results, one can characterize the steady state of
the motion. For instance, the measurement of the linewidth of the motional
sidebands gives the cooling rate of the process. The phonon number $\langle
n\rangle$ at steady state can be measured through the ratio between the
heights at the center-frequency of the motional sideband and of one
peak of the incoherent spectrum.
In this way, one gets a simple 
equation at second order in $\langle n\rangle$ whose coefficients are determined only by the laser parameters. We remark that for larger values of 
$\langle n\rangle$ the visibility of the motional sidebands over the Mollow--type inelastic spectrum increases. This is illustrated in the inset of Fig.~4.

In summary, the spectrum we have evaluated allows
to gain insight into the quantum dynamics and steady state of the
atom interacting with light. Our results
are in agreement with the dynamical picture presented in
section II. This picture is based  on a clear separation
between the two time scales $T_0$, $T_1$, on which also 
the validity of the perturbative expansion lies. Analytical 
estimates and numerical checks of the validity of this coarse--grained
dynamics have been presented in~\cite{Giovanna03}.

It is interesting to compare these results with the features found in the
emission spectrum of a trapped two-level atom. 
In a two-level transition driven by a
plane wave, the incoherent spectrum has a contribution 
at zero order in the Lamb--Dicke
expansion, as at this order the internal steady state of the system is characterized by
non-vanishing occupation of the excited state. The features due
to the mechanical effects manifest here in the motional sidebands. These are
narrow resonances, whose width is the cooling rate. However, at a fixed
detection angle $\psi$ the curves are Fano-like profiles, whose
relative height varies with $\psi$ (while, once integrated over the
solid angle of emission, have Lorentz shape and are
equal)~\cite{Lindberg86,Cirac93}. 
This behaviour
is an interference effect between Raman processes at
second order in the Lamb--Dicke expansion \cite{Cirac93}: Given $|g\rangle$,
$|e\rangle$ ground and excited states of the dipole transition, the Raman
processes $|g,n\rangle\to |e,n\rangle\to |g,n\pm 1\rangle$ and $|g,n\rangle\to
|e,n\pm 1\rangle\to |g,n\pm 1\rangle$ are of the same order and lead to the
emission of the photon. They therefore interfere, and their interference
signal (the height of the sidebands) is modulated by the emission angle. 

This
behaviour disappears when the dipole is at the node of a standing wave:
Then, the carrier transition $|g,n\rangle\to |e,n\rangle$
is suppressed and at this order only the transitions $|g,n\rangle\to |e,n\pm
1\rangle\to |g,n\pm 1\rangle$ occur. Hence, 
both motional sidebands are Lorentz curves of equal
shape, independently of the emission angle (which just affects the total
height of the signal, according to the dipole pattern of radiation).
Thus, in this case transitions in zero order in the Lamb--Dicke expansion
vanish because of the spatial mode structure, while light scattering 
occurs because of the spatial gradient of the field intensity over the
center--of--mass wave packet. Remarkably, in our case transitions in zero order
in the Lamb--Dicke expansion are suppressed because of quantum interference
between dipole excitation paths, and photon scattering occurs due to the phase
gradient of the field over the center--of--mass wave packet. 

\section{Conclusions} 

\noindent 
We have presented a theoretical study of the spectrum of fluorescence of a
trapped atom whose internal degrees of freedom are driven in a 
$\Lambda$--configuration at two photon resonance. In this system, the atomic
emission at  
steady state is only due to the mechanical effects of the atom--photon
interaction. The spectrum has been evaluated at
second order in the Lamb--Dicke expansion, i.e.\ in the expansion of the size
of the atomic wave packet over the laser wavelength. We find that the
spectrum is
characterized by two narrow resonances corresponding to the motional
sidebands, i.e.\ the Stokes and anti-Stokes components, 
and by a Mollow--type inelastic spectrum, while the elastic peak is at higher
order. Through these properties, several
information about the quantum dynamics and steady state
of the driven atom can be
extracted, like the cooling rate and the temperature, and the contributions 
of the individual
scattering processes can be identified. Furthermore, for
relatively large 
temperatures the sidebands of the elastic
peak may be orders of magnitude larger than any other spectral signal, and
the spectrum can be said to be solely composed of these two frequencies. 

Our results provide an interesting insight into the underlying physics of
the mechanical effects of light-atom interaction, and may contribute to 
on--going experiments investigating and engineering 
the coupling of single trapped atoms and ions with
electromagnetic fields.

\section{Acknowledgements}

\noindent 
The authors are grateful to
Wolfgang Schleich for stimulating discussions and helpful comments.
G.M.\ ackowledges several clarifying discussions with J\"urgen Eschner.

\begin{appendix}

\section{Perturbation Theory}

\noindent 
The equations (\ref{Sec:Pert:1}) to solve iteratively 
in the perturbative expansion are 
\begin{eqnarray}
  \label{Eq:rho:1} &&{\cal L}_0\rho_1^{\lambda}+{\cal
  L}_1\rho_0^{\lambda}=\lambda_0\rho_1^{\lambda}+\lambda_1\rho_0^{\lambda}, \\
  &&{\cal L}_0\rho_2^{\lambda}+{\cal L}_1\rho_1^{\lambda}+{\cal
  L}_2\rho_0^{\lambda}=\lambda_0\rho_2^{\lambda}
  +\lambda_1\rho_1^{\lambda}+\lambda_2\rho_0^{\lambda}, \label{Eq:rho:2}
  \end{eqnarray} where  $\rho_0^{\lambda}$ satisfy ${\cal
  L}_0\rho_0^{\lambda}=\lambda_0\rho_0^{\lambda}$. For
 $\lambda_0=0$, $\rho_0=\rho_{\rm D}\mu$, with $\mu$ given in Eq.\
  (\ref{mu}). Equation (\ref{Eq:rho:1}) gives 
\begin{equation} \label{rho:1}
  (1-{\cal P}_0^{\lambda})\rho_1^{\lambda}=-\frac{1-{\cal P}_0^{\lambda}}
  {\lambda_0-{\cal L}_0}(\lambda_1-{\cal L}_1)\rho_0^{\lambda}, 
\end{equation}
  where ${\cal P}_0^{\lambda}$ is the zero-order projector onto the subspace
  at eigenvalue $\lambda$, ${\cal P}_0^{\lambda}=\rho_0^{\lambda}\otimes
  \check{\rho}_0^{\lambda}$. Inserting (\ref{rho:1}) in (\ref{Eq:rho:2}) we obtain
  \begin{eqnarray} \label{rho:2} &&(1-{\cal
  P}_0^{\lambda})\rho_2^{\lambda}=-\frac{1-{\cal P}_0^{(0)}}{\lambda_0-{\cal
  L}_0} \\
&&\times \Bigl[-(\lambda_1-{\cal L}_1)\frac{1-{\cal
  P}_0^{\lambda}}{\lambda_0-{\cal L}_0}(\lambda_1-{\cal L}_1)
+(\lambda_2-{\cal L}_2)\Bigr]\rho_0^{\lambda}. \nonumber
\end{eqnarray} Analogously,
  one finds the perturbative corrections to the left eigenelements
  $\check{\rho}_0^{\lambda}$ solving equations (\ref{Sec:Pert:2}) at second
  order. This in turn allows to evaluate the perturbative
  corrections to the projectors ${\cal P}^{\lambda}_0$. Using ${\cal
  P}_1^{\lambda}= \rho_0^{\lambda}\otimes \check{\rho}_1^{\lambda}+
  \rho_1^{\lambda}\otimes \check{\rho}_0^{\lambda}$, we obtain~\cite{Cirac93}
\begin{eqnarray}
&&(1-{\cal P}_0^{\lambda}){\cal P}_1^{\lambda}=\frac{1-{\cal
  P}_0^{\lambda}}{\lambda_0-{\cal L}_0}{\cal L}_1{\cal P}_0^{\lambda},\\
  &&{\cal P}_1^{\lambda}(1-{\cal P}_0^{\lambda})={\cal P}_0^{\lambda}{\cal
  L}_1\frac{1-{\cal P}_0^{\lambda}}{\lambda_0-{\cal L}_0}. \end{eqnarray}
The equations for the
  corrections $\lambda_1$, $\lambda_2$ to $\lambda_0$ are found by multiplying
  Eqs.\ (\ref{Eq:rho:1}), (\ref{Eq:rho:2}) by
  $\check{\rho}_0^{\lambda}$ on the left
and taking the trace. The resulting equations are
  \begin{eqnarray} \label{lambda:1} &&\lambda_1={\rm
  Tr}\{\check{\rho}^{\lambda}_0{\cal L}_1\rho_0^{\lambda}\}=0,
\\ &&\lambda_2={\rm
  Tr}\{\check{\rho}^{\lambda}_0{\cal L}_2\rho_0^{\lambda}\}- {\rm
  Tr}\{\check{\rho}^{\lambda}_0(\lambda_1-{\cal
  L}_1)\rho_1^{\lambda}\}\label{lambda:2},\\ &&={\rm
  Tr}\{\check{\rho}^{\lambda}_0{\cal L}_2\rho_0^{\lambda}\}
+{\rm Tr}\{\check{\rho}^{\lambda}_0{\cal L}_1\frac{1-{\cal
  P}_0^{\lambda}} {\lambda_0-{\cal L}_0}{\cal L}_1)\rho_0^{\lambda}\}\nonumber,
\end{eqnarray} where we have used Eq.\ (\ref{rho:1}) and
  relation $\check{\rho}_0^{\lambda}{\cal
  L}_0=\lambda_0\check{\rho}_0^{\lambda}$. From Eq.\ (\ref{lambda:1}) it is
  visible that $\lambda_1=0$ for all eigenvalues $\lambda$: In fact, the
  Liovillian ${\cal L}_1$ couples subspaces at different $\lambda_E$, but it
  vanishes inside a subspace at fixed $\lambda_E$, ${\cal P}_0^{\lambda}{\cal
  L}_1 {\cal P}_0^{\lambda}=0$.

\section{Contributions to the spectrum in second order}

\noindent 
The term at zero order in the perturbative expansion
    \begin{eqnarray*} 
S_0(\omega)=\sum_{\lambda_0}\frac{1}{{\rm i}(\omega-\omega_{L,1})-\lambda_0}
    {\rm Tr} \left\{D_0^{\dagger}{\cal P}_0^{\lambda}D_0\rho_0\right\}=0
    \end{eqnarray*} vanishes, since 
\begin{equation}
\label{Drho}
D\rho_0=\rho_0D^{\dagger}=0,
\end{equation}
as there is no
excited state occupation at steady state in zero order. Analogously, in
first order it can be shown that
\begin{eqnarray*}
    S_1(\omega)&=&\sum_{\lambda_0}
\frac{1}{{\rm i}(\omega-\omega_{L,1})-\lambda_0}\Bigl({\rm
    Tr}\left\{D_0^{\dagger}{\cal P}_1^{\lambda}D_0\rho_0\right\}\\ &+& {\rm
    Tr}\left\{D_1^{\dagger}{\cal P}_0^{\lambda}D_0\rho_0\right\}+ {\rm
    Tr}\left\{D_0^{\dagger}{\cal P}_0^{\lambda}D_1\rho_0\right\}\\ &+& {\rm
    Tr} \left\{D_0^{\dagger}{\cal
    P}_0^{\lambda}D_0\rho_1\right\}\Bigr)=0,\end{eqnarray*} where each of 
the first three terms are equal to zero because of relation~(\ref{Drho}). The last
term is equal to zero because here the position operator $\hat x$ occurs
linearly.

The non-vanishing contributions to the spectrum at second order
are shown in Eq.\ (\ref{S:4}). All other terms vanish. For most of them,
this can be demonstrated using (\ref{Drho}). We would like to emphasize the 
disappearance of the contributions 
\begin{eqnarray*}
&&{\rm Tr}\{D_1^\dagger{\cal
    P}_1^{\lambda}D_0 \rho_0\}={\rm  Tr}\{D_0^{\dagger}{\cal P}_1^{\lambda}D_1
\rho_0\} =0,\\
&&{\rm Tr}\{D_1^\dagger{\cal
    P}_0^{\lambda}D_0 \rho_1\}={\rm  Tr}\{D_0^{\dagger}{\cal P}_0^{\lambda}D_1
\rho_1\} =0,\\
&&{\rm Tr}\{D_2^\dagger{\cal
    P}_0^{\lambda}D_0 \rho_0\}={\rm  Tr}\{D_0^{\dagger}{\cal P}_0^{\lambda}D_2
\rho_0\} =0,\\
\end{eqnarray*}
which, together with ${\cal K}\rho_0=0$, imply that the spectral signal does
not depend on the angle of emission up to second order.

\end{appendix}


\begin{thebibliography}{11}
\bibitem{Footnote}
Here we assume that the frequencies $\omega_1$, $\omega_2$ are sufficiently
different such that at a certain bandwidth the detector measures the resonance
fluorescence of the light scattered by one dipole transition. This assumption
simplifies the notations, and the treatment suffers no loss of generality.
\bibitem{Footnote:mu}
The density matrix $\mu$ is the stationary solution of the master
equation obtained by adiabatically eliminating the internal degrees of 
freedom at second order in the Lamb--Dicke parameter~\protect\cite{Giovanna03}.
\bibitem{FootnoteN}
In particular, the spectral signal scales at all orders with 
$\eta\propto k_1\cos\phi_1-k_2\cos\phi_2$, such that it disappears 
for $k_1\cos\phi_1=k_2\cos\phi_2$, where at steady state there is no photon
emission.
\end{thebibliography}
\end{document}